\documentclass[pre,epsfig,aps,twocolumn]{revtex4}

\usepackage{epsfig}
\usepackage{fancyheadings}
\usepackage{amstext}

\begin{document}

\sloppy
\pagestyle{empty}

\title{Inferring DNA sequences from mechanical unzipping: an
  ideal-case study}
\author{V. Baldazzi $^{1,2,3}$, S. Cocco $^2$, E. Marinari $^4$, 
R. Monasson $^3$}
\affiliation{
$^1$ Dipartimento di Fisica, Universit\`a di Roma
{\em Tor Vergata}, Roma, Italy\\
$^2$ CNRS-Laboratoire de Physique Statistique de l'ENS, 24 rue Lhomond,
75005 Paris, France\\
$^3$ CNRS-Laboratoire de Physique Th\'eorique de l'ENS, 24 rue Lhomond,
75005 Paris, France\\
$^4$ Dipartimento di Fisica and INFN, Universit\`a di Roma 
{\em La Sapienza}, P.le Aldo Moro 2,  00185 Roma, Italy
}

\begin{abstract}
We introduce and test  a method to predict the
sequence of DNA molecules from {\em in silico}  unzipping
experiments.  The method is based on Bayesian inference
and on the Viterbi decoding algorithm.  The probability of misprediction 
decreases exponentially with the number of unzippings, with a decay 
rate depending on the applied force and the sequence content.
\end{abstract}

\maketitle


DNA molecules are the support for the genetic information, and
knowledge of their sequences is very important from the  biological and
medical points of view. State-of-the-art DNA sequencing methods  
rely on biochemical and gel electrophoresis techniques \cite{mb}, and
are able to correctly predict about 99.9\% of the bases. They were massively 
used over the past ten year to obtain the human genome (and the ones
of other organisms).  

\begin{figure}
\begin{center}
\psfig{figure=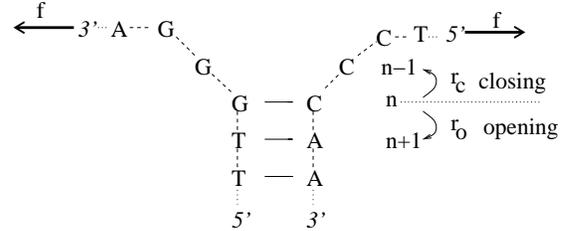,height=3cm,angle=0}
\end{center}
\caption{An unzipping experiment. The extremities of the molecule are 
stretched apart under a force $f$. The fork at location $n$ (nb. of
open base pairs) moves 
backward or forward with rates (probability per unit of time)
$r_c$  and $r_o$ (\ref{ratemd}).}
\label{fig1}
\end{figure}

\begin{figure}
\begin{center}
\psfig{figure=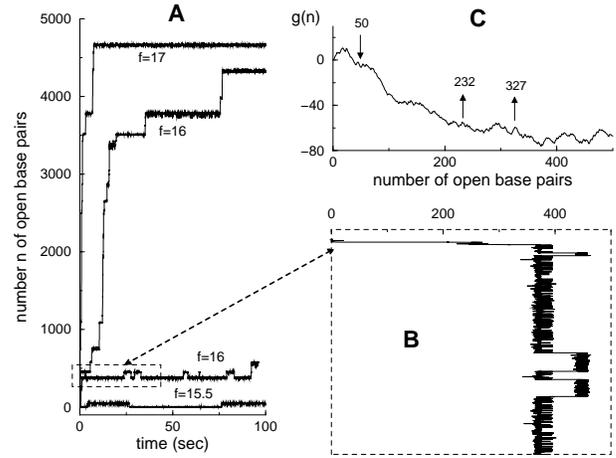,height=8cm,angle=-90}
\end{center}
\caption{Fixed-force unzipping of $\lambda$-phage. {\bf A.} number $n$ of
open base pairs vs. time $t$ for forces $f$ ranging from $15.5$ to
$17$ pN from  model (\ref{ratemd}). {\bf B.}
magnification of the boxed region in {\bf A} 
after a $90$ degree clockwise rotation. {\bf C.} free
energy landscape $g(n)$ versus $n$ for the first $450$ bases and
$f=16$ pN.  Down and up arrows indicate, respectively, a local minimum
in $n=50$ and two maxima in $n=232$ and $n=327$ (see the text).}
\label{fig2}
\end{figure}

Nevertheless, the quest for alternative (cheaper and/or faster) 
sequencing methods is an active field of research. In this
regard, recent single molecule micro-manipulations are of particular
interest. Among them are DNA unzipping under a mechanical action
\cite{Ess97,Boc02,Lip,Har03,Dan03}
or due to translocation through nanopores
\cite{Mat04}, the observation of the sequence-dependent
activity of an exonuclease \cite{Van03,Per03}, the optical analysis of
DNA polymerization in a nano-chip device \cite{Lev03}, the detection of
single DNA hybridization \cite{Zoc03}.  Hereafter, we focus on
mechanical unzipping (see Figure~\ref{fig1}), first realized by
Bockelmann, Heslot and coworkers in 1997 \cite{Ess97,Boc02}. In their
experiment, the strands are pulled apart under a constant velocity.
The force is measured and fluctuates around $15$ pN for
the $\lambda$-phage DNA (a $48,502$ base long virus), with higher
(respectively, lower) values corresponding to the unzipping of GC (AT)
rich regions.  Researchers have also unzipped RNA
molecules \cite{Mat04,Lip,Har03}, or DNA under a constant force
(instead of velocity) \cite{Dan03}.  Figure~\ref{fig2}A sketches a
fixed-force output signal, with its pauses in the opening at
sequence-specific positions.

Various theoretical works have studied and reproduced 
the unzipping signal related to a given sequence
\cite{Boc02,Coc3,Coc4,Lub,Hwa,Felix,mar}. Hereafter we address
the inverse problem: given an unzipping signal (for example the one of
Figure \ref{fig2}A), can we predict the underlying sequence?  We
propose a Bayesian inference method to solve this problem
\cite{bayes},  and test
it {\em in silico} on the $\lambda$-phage.  We analytically study the
dependence of the quality of the prediction on the sequence content,
on the force, and on the number of unzippings.
Finally we list the main obstacles to be circumvented prior
to practical applications.

Let ${\cal S}=\{b_1,b_2,\ldots,b_N\}$ denote the sequence of $N$ bases
along the $5'\to3'$ strand (the other strand is
complementary).  We model the unzipping of the molecule through the
evolution of the number $n$ of open base pairs \cite{Coc4};  
base pair opening ($n\to n+1$) and closing ($n\to n-1$) happen with rates
(Figure~\ref{fig1})
\begin{equation}
r_o (n) = r\; \exp\{g_0(n)\} \; ,  \;\; 
r_c     = r\; \exp\{g_{ss}\} \; .
\label{ratemd}
\end{equation}
$g_0(n)$ is the binding energy of base pair (bp) $n$ in units
of k$_B$T \cite{Zuk}; 
it depends on the base $b_n=A,T,G$, or $C$ and,
due to stacking effects, on the nearest base $b_{n+1}$.
$g_{ss}$ is the work needed to stretch an open bp under a force $f$
in units of k$_B$T ; 
according to the modified freely--jointed--chain model 
\cite{Coc3}, ${g}_{ss}= -2 
\ell/\ell_0\,\ln [\sinh(x)/x]$ where $x\equiv \ell_0 \, f / k_B T$,
and $\ell_0=15$ {\AA} and $\ell= 5.6$ {\AA} are, respectively, the Kuhn
and  effective nucleotide lengths.  
Relation (\ref{ratemd}) implies that the opening rate
at base $n$ is a function of the sequence, $r_o(n)=r_o(b_n,b_{n+1})$,
while the closing rate $r_c$ only depends on the force \cite{lungo}.
This {\em a priori} choice has been shown \cite{Coc4} to reproduce
quantitatively the behavior of unzipping experiments on short
polynucleotides \cite{Lip}, with a typical frequency $r \simeq
10^{6-7}$ sec$^{-1}$.

Rates (\ref{ratemd}) define a one-dimensional biased random walk for
the fork position (number of open bp) $n(t)$ in the potential
$\displaystyle{g(n)=n \, g_{ss} -\sum_{i=1} ^n g_0(i)}$, that can be
interpreted as the free energy of the molecule when the first $n$ bp
are open.  We show in Figure~\ref{fig2}B\&C a typical time-trace of $n(t)$
generated by Monte Carlo (MC) simulation for the $\lambda$-phage sequence,
together with the free energy landscape $g(n)$. Plateaus of $n(t)$
coincide with deep local minima of $g(n)$, where the fork remains
trapped for a long time. As the force increases, opening becomes more
favorable, and plateaus shrink.

Our {\em in silico} time-traces are stochastic due to the thermal noise: 
two runs will give different traces. The probability of a time-trace
only depends on the set ${\cal N}=\{t_n,u_n,d_n\}$
of times $t_n$ spent on each base $n$, and of numbers  $u_n$ and  
$d_n$ of up ($n\to n+1$) and down ($n\to n-1$) transitions respectively.
Given the sequence ${\cal S}$, 
this probability reads
\begin{equation}
{\cal P}({\cal N} | {\cal S} )= c \prod_n \,M (b_n,b_{n+1}; t_n,u_n,d_n)\;,
\label{p}
\end{equation}
where $c$ is a (sequence-independent) normalization constant and
$M (b_n,b_{n+1} ;t_n,u_n,d_n) =r_o\left(b_n,b_{n+1}\right)^{u_n} \,
r_c^{d_n}\; \exp\{-(r_o(b_n,b_{n+1})+r_c)t_n \}$.
Equation (\ref{p}) provides the solution of the direct problem:
given the sequence ${\cal S}$ what is the distribution of the 
time-traces ${\cal N}$? The inverse problem, that is the prediction
of the sequence given some time-trace, can be
addressed within the Bayesian inference framework.
The probability that DNA sequence is ${\cal S}$  given an observed ${\cal N}$
is \cite{bayes}
\begin{equation} 
\label{bayes}
{\cal P}({\cal S}|{\cal N})= \frac{{\cal P }( {\cal N }|{\cal S}) 
\;{\cal P}_0({\cal S}) }{ {{\cal P}({\cal N})}}\;.
\end{equation}
The value of ${\cal S}$ that maximizes this probability, ${\cal S}^*$,
is our prediction for the sequence. In the absence of any {\em a
priori} information about the sequence, ${\cal P}_0({\cal S})$ is the flat
distribution, equal to $4^{-N}$. The maximization of ${\cal P}({\cal
S}|{\cal N})$ then reduces to that of
${\cal P}( {\cal N}|{\cal S})$  (\ref{p}).

In practice the most likely sequence ${\cal S}^*$ may be found using
the Viterbi algorithm \cite{viterbi}. The procedure is equivalent to
a zero temperature transfer matrix technique exploiting the
nearest-neighbor nature of couplings between bases in (\ref{p}). The
probability $P_n$ for the base $b_{n}$ fulfills the recursive equation
\begin{equation} 
\label{recur}
P_{n+1}(b_{n+1}) \propto \max_{b_{n}} \; P_n(b_n) \, M (b_{n},b_{n+1} ;
t_n,u_n,d_n) \;,
\end{equation}
where the proportionality constant is irrelevant for our purpose.  The
maximum in (\ref{recur}) is reached for some base $b_n^{max}
(b_{n+1})$ that depends on the next base $b_{n+1}$.  Starting 
from $P_1 (b_1)= \frac 14$, we obtain the probability
$P_N(b_N)$ for the last base of the sequence through iterations of
(\ref{recur}). Maximization of $P_N(b_N)$ yields the most likely value
for this last base, $b^*_N$.  The whole optimal sequence ${\cal S}^*$
is then recursively obtained from the relation $b^*_{n-1} =
b_{n-1}^{max} (b_n^*)$.

\begin{figure}
\begin{center}
\psfig{figure=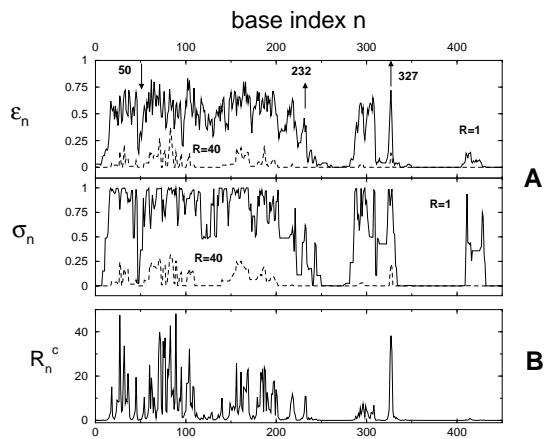,height=7cm,angle=-90}
\caption{{\bf A}. Probability $\epsilon _n$ of an error (top) 
and entropy $\sigma _n$ (middle) versus the base index $n$, 
for the first
$450$ bp of DNA $\lambda$-phage at $f=16$~pN. Full lines correspond
to $R=1$ unzipping, dotted lines to $R=40$. 
{\bf B.} Theoretical values for the  decay constants $R_n^c$  in
 $\epsilon _n$  (\ref{rcf}). For instance, base
$232$ (arrow) is characterized by $R_{232}^c \simeq 10$, and
is not (respectively, well) predicted with $R=1$ (resp. $R=40$)
unzippings. }
\label{fig3}
\end{center}
\end{figure}
\begin{figure}
\begin{center}
\psfig{figure=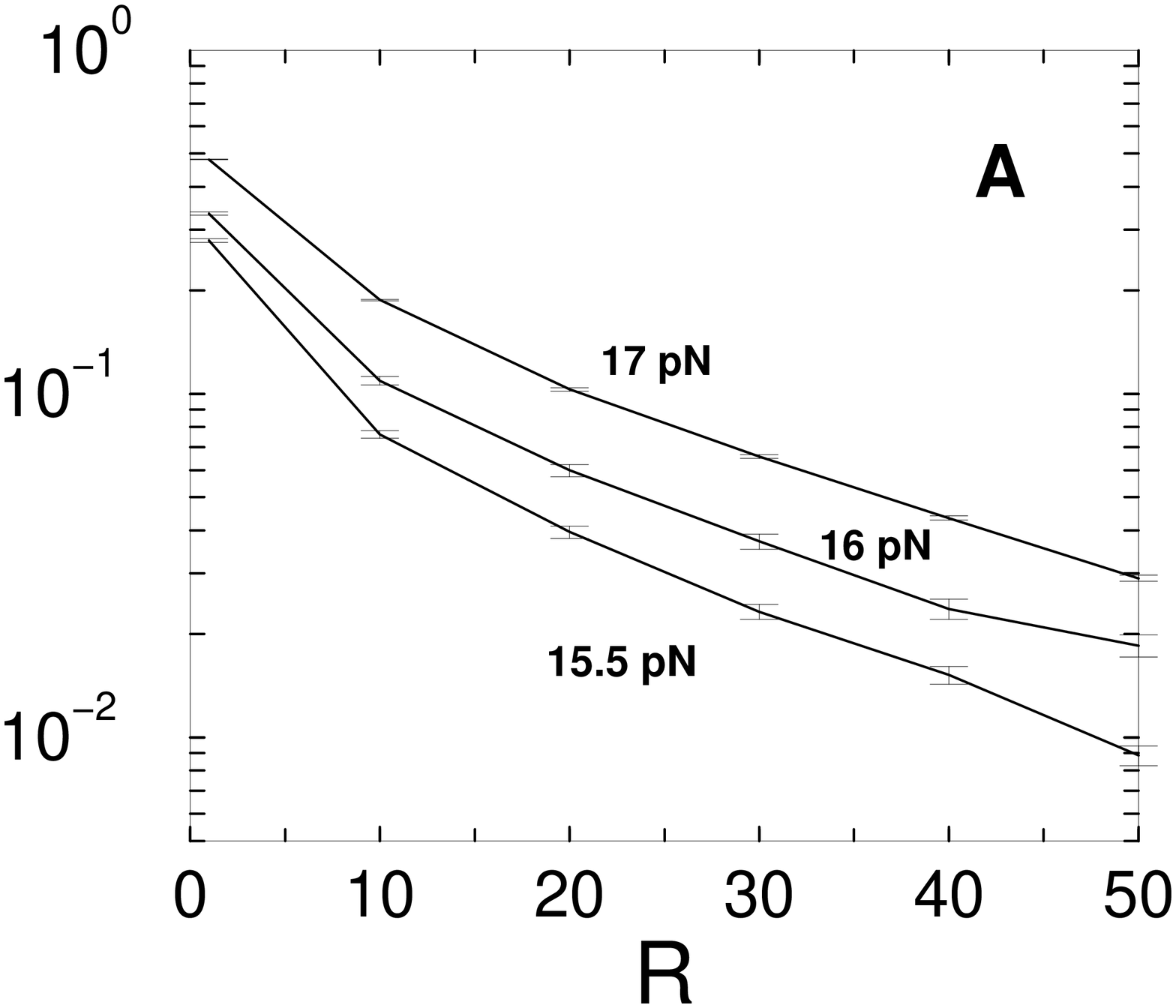,height=4cm,angle=-90}
\psfig{figure=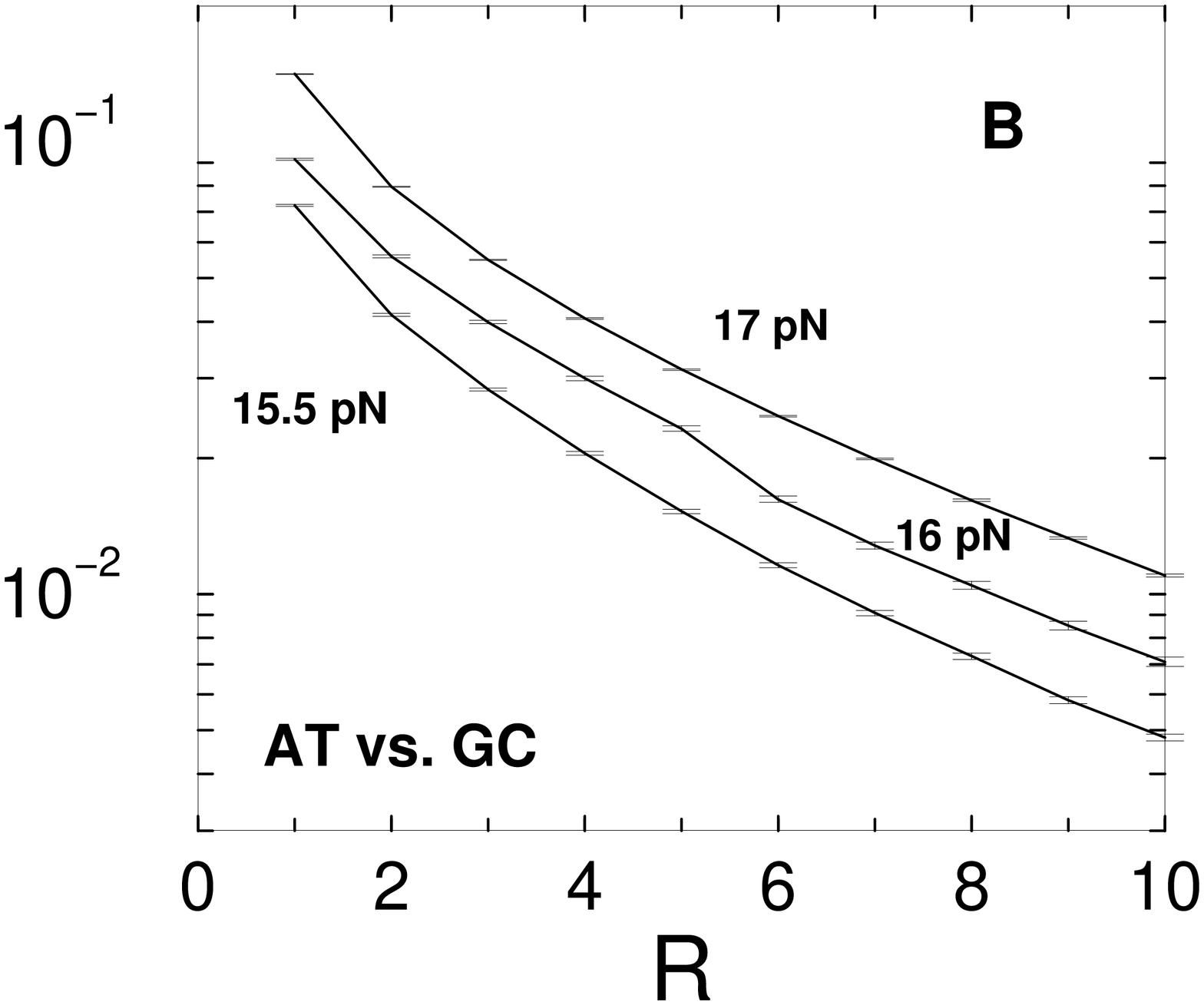,height=4cm,angle=-90}
\caption{{\bf A.}
Fraction $\epsilon$ of mispredicted bases for the $\lambda$-phage
versus the number $R$ of unzippings, averaged over $1000$ samples of
$R$ unzippings, and
for forces of $15.5$, $16$ and $17$ pN (from bottom to top).
{\bf B}. Same as {\bf A}, but we only discriminate among weak and
strong basis.}
\label{fig4}
\end{center}
\end{figure}

We have tested our sequencing method on the $\lambda$--phage.  First
we build a dynamical process on the sequence ${\cal S}^\lambda$ of the phage 
with rates (\ref{ratemd}), and 
generate an unzipping trace ${\cal N}$ by a MC procedure.  Then we use
the Viterbi procedure (which ignores the phage sequence) to make a
prediction for the sequence, ${\cal S}^*$, from this signal ${\cal
N}$. We estimate the error over the prediction about base $n$ from
the failure rate
\begin{equation} \label{defom}
\epsilon _n = \hbox{\rm Probability} 
\left[ b_n^*  \ne b_n^\lambda \right]\;,
\end{equation}
where the probability is computed by repeating the procedure over
different MC runs. 
The errors $\epsilon _n$ are shown in Figure~\ref{fig3} (with
the continuous curve) for the first $450$ bases 
at a force of $16$ pN.  Values range from $0$ (perfect prediction) to
$0.75$ (random guess of one among four bases).  A comparison with the
free energy $g(n)$ (Figure~\ref{fig2}) shows that $\epsilon _n$ is
small in the flattest part of the landscape ($350< n< 450$), or in
local minima e.g. the $n=50$ base
preceded by 4 weak bases and followed by 4 strong bases
(...TTTA-A-GGCG...).  Conversely, bases that are not well determined
correspond to local maxima of the landscape e.g. $n=327$, $328$ bases
between $7$ strong and $7$ weak bases
(...GCCGCCG-TC-ATAAAAT...).  We plot the average fraction of
mispredicted bases, $\displaystyle{\epsilon = \frac 1 N \sum_{n}
\epsilon _n}$, in Figure~\ref{fig4}A.  As shown in Fig.~\ref{fig2},
for a larger force, there are more open bases (about $60$, $600$ and
$5000$ at $15.5$, $16$ and $17$ pN in about $100$ seconds), but the
time spent on each base is smaller, and therefore $\epsilon$ is larger
($\epsilon =20 \%,23\%,47\% $).
Most errors are due to the difficulty of distinguishing A from T, and
G from C. The probability  that a weak
(A or T) base is confused with a strong one (G or
C), or vice-versa, is plotted in Figure~\ref{fig4}B.

Performances can be greatly improved by collecting information from
multiple unzippings. As the
number of passages over the same base $n$ gets larger, the total
waiting times $t_n$ and transition parameters $u_n,d_n$ become less
affected by fluctuations, and reflect more faithfully the
thermodynamic signature of the base. In practice, we look for the 
most likely sequence ${\cal S}^*$ given $R$ unzipping signals
${\cal N}_1, {\cal N}_2,\ldots , {\cal N}_R$. 
Figures \ref{fig3}A and \ref{fig4} shows the drop down in the 
probability of error when the number $R$ of 
unzippings increases. Observe from Figure \ref{fig3}A that the decay of 
$\epsilon _n$ with $R$ (\ref{defom}) varies from base to base.
The decrease of the total error $\epsilon$ 
is much faster for AT vs. GC (Figure~\ref{fig4}B) than for 
complete (Figure~\ref{fig4}A) recognition. 

It is useful to build indicators of performances that do not rely on
the exact knowledge of the unzipped sequence (used here for checking
the quality of our results but unknown in practical
applications). To this aim, we calculate the optimal sequences $S^*_b$
when base $n$ is constrained to value $b$, and the corresponding
probabilities $P_n^*(b)$. 
We then define the Shannon entropy
\begin{equation}
\sigma_n=-  \sum_{b=A,T,G,C}\, \langle P_n^*(b)\,
\log _4 P_n^*(b) \rangle\; ,
\end{equation}
where $\langle\cdot\rangle$ denotes the average over MC data.  $\sigma
_n$ is low when one of the four bases has much higher probability than
the other ones and close to unity for uncertain predictions
(equiprobable bases).  Figure \ref{fig3} shows that $\sigma _n$ and
$\epsilon_n$ as a function of the base index $n$ are indeed very
similar: the Shannon entropy is a good indicator of the success of our
reconstruction.


Our analytical study of the dependence of the quality of 
the prediction upon the force, the sequence content, and the number 
of unzippings confirms that the probability of error $\epsilon _n$
decreases very quickly with $R$,
\begin{equation} \label{rcf}
\epsilon _n \sim e^{ -R/R^c_n} \ . 
\end{equation}
 As $f$ decreases to its critical value (below which the
molecule cannot open), the decay constant
$R_n^c$ decreases to zero, and predictions 
drastically improve at fixed $R$.
Our theoretical values for $R^c _n$ are shown in Figure~\ref{fig3}B
for $f=16$ pN, and vary from 0.1 to 45 with the base index $n$. The
agreement with the decay of $\epsilon _n$ from $R=1$ to $40$ unzippings
(Figure \ref{fig3}A) is excellent. Note that $\epsilon$ in
Figure~\ref{fig4} is not a pure exponential, but a superposition of
exponentials with $n$-dependent decay constants $R^c_n$.  
We now present the calculation of $R^c_n$ in three steps.

{\em (a) Pairing only, high force.} 
Assume first that there are only 2 and not 4 bp-types, called
$+$ and $-$, and no stacking interaction. Call $\Delta$ the difference 
between the (pairing) free-energies of $+$ and $-$ bp, and  $\langle
t_\pm\rangle$ the average time spent by the
fork on a $\pm$ bp before moving forward or backward. Consider
now a bp of type $b$ and call $t$ the time spent on this bp
divided by the number $R$ of unzippings. From the central limit
theorem, for large  $R$, 
$t$ gets narrowly peaked around its mean value $\langle
t_b\rangle$, with Gaussian fluctuations $\delta t\sim R^{-\frac 12}$. 
Bayes prediction (\ref{bayes}) will be erroneous,
$b^*=-b$, when  $t$ is closer to $\langle
t_{-b}\rangle$ than to its expected value $\langle t_b\rangle$.  
The probability of error is thus given
by the Gaussian tail, and scales as $\epsilon \sim \exp( - \delta t^{-2})$,
hence (\ref{rcf}).  A careful calculation \cite{lungo} gives the
precise value of the decay constant in (\ref{rcf}),
\begin{equation}\label{nostackomega}
R ^c= \frac 1{\tau -1 -\ln \tau}\quad \hbox{\rm with}
\quad \tau = \frac {\Delta}{1- e^{-\Delta}} \ .
\end{equation}
Good predictions are obtained when  the molecule is unzipped a 
few $R^c$ times (for example $R \simeq 4 R^c$ 
gives $\epsilon \simeq 2\%$). 
To distinguish weak (AT) from strong (CG) bp only we have
$\Delta \simeq 2.8$ \cite{Zuk} and  
$R^c\simeq 1$ (Figure~\ref{fig4}B), while complete recognition corresponds
to $\Delta \simeq 0.5$ and  $R^c \simeq 30$ (Figure~\ref{fig4}A). 

{\em (b) Pairing and Stacking, high force.} In presence
of stacking interactions, the error $\epsilon _b$ on base $b$
depends on the neighboring bases, say, $x$ and $y$. 
 At large $R$, errors are rare and
are typically due to a single base mis-prediction e.g. $b\to b'$. The
probability $\epsilon_{b\to b'}$ of this mistake is the product of the
probabilities $\epsilon _{xb\to xb'}$ and $\epsilon _{by\to b'y}$ of
the two bond violations. We estimate $ \epsilon
_{xb\to xb'} \sim e^{-R/R^c_{xb\to xb'}}$ 
from (\ref{rcf}) where $R^c_{xb\to xb'}$ is
given by (\ref{nostackomega}) with $\Delta = g_0^{xb'}-g_0^{xb}$.
A similar expression is readily obtained for the $by$ bond. Knowing
the asymptotic behavior of $\epsilon _{b\to b'}$, we calculate 
$\epsilon _b \sim e^{-R/R^c_{xby}}$ by selecting the worst value for $b'$, 
\begin{equation} 
\label{id}
\frac 1{R^c _{xby}} = \min _{b' (\ne b)} \left[  \frac 1{R^c _{xb \to xb'}}
+ \frac 1{R^c _{by \to b'y}} \right] \ .
\end{equation}
The above derivation is confirmed by exact calculations based on
techniques for 1D disordered systems \cite{diso,lungo}.

{\em (c) Moderate force.}
The above calculations are correct for high forces. At moderate forces,  
bp can close and are visited several times by the fork. The effective number 
of unzippings is $R\times \langle u_{n}\rangle$, where $\langle u_n\rangle$  
is the average number of openings of bp $n$ during a single unzipping. 
The decay constant is thus, from (\ref{rcf}),
\begin{equation} 
R^c_n =  {R ^ c _{b_{n-1}b_nb_{n+1}} }/{\langle u_n\rangle }\ .
\end{equation}
As the force is lowered, $\langle u_n\rangle$ increases (from 1 
at high force), and $R^c_n$ diminishes.  To
calculate $\langle u_n\rangle$, we consider the 1D transient random walk 
defined by the probabilities $q_m\equiv
r_c/(r_o(m)+r_c)$ and $1-q_m$ for closing or opening bp $m$. 
Let $p_{m}^{(n)}$ be the probability that the fork will never
reach position $n$ starting from $m (>n)$.  The ratios $\rho _m^{(n)}
= p_{m}^{(n)}/p_{m+1}^{(n)}$ fulfill the Riccati recursion relation
\cite{lungo} $\rho _{m+1} ^{(n)} = (1- q_{m+1}) / (1-q_{m+1} \, \rho
_m ^{(n)} )$.  Iterating with boundary condition $\rho_n^{(n)}=0$
allows us to obtain $\langle u _ n\rangle = 1/p_n^{(n+1)} = \prod
_{m>n} \rho _m^{(n)}$. 


Finally we discuss the difficulties hindering a direct application of
our inference method to real data (see also \cite{hwa}), 
and possible way-outs. 

First, temporal resolution is limited in practice. The frequency bandwidth
is controlled by the viscous friction and the stiffness of the
setup, with a typical value of $10$ kHz \cite{Boc02,bustt}. The
corresponding time, $\delta\tau \simeq 100$ $\mu$sec, is about $10$
(resp. $200$) times longer than the typical opening time for GC
(resp. AT) bp.  As a result, the fork can move by $D (> 1)$ bp during
the time interval $\delta\tau$.  We have taken into account such moves by
considering interactions between bases at distance $\le D$ in the
probability $P(\cal N|S)$, and modified the
reconstruction procedure accordingly (the transfer matrix has now
dimension $4^D$) \cite{lungo}.  In practice, when
$\delta\tau = 1\ \mu$sec, sequences cannot be predicted with the usual
$D=1$ reconstruction procedure, but are correctly inferred with the
$D=6$ procedure.  Though time resolution is currently far below this
limit, future experimental progresses, and new technologies e.g.
combination of optical trap and single-molecule fluorescence
\cite{Lan03}, could help bridging the gap.

Secondly, thermal fluctuations 
of the open strands lead to  an uncertainty $\delta n$ 
over the position $n$ of the fork \cite{siggia} e.g. $\delta n \simeq 5$
for $f\simeq 15$ pN and $n=300$ open bp \cite{Coc3}. The presence
of correlations between bases at distance $D\le \delta n$ does not
affect the result (\ref{rcf}) for $\epsilon _n$  as long as the relaxation
time of the strands is smaller than the bp opening time {\em
i.e.} up to a few hundreds open bp. What happens for larger values of $n$
is currently under study. 

Thirdly, we have assumed so far to have a perfect knowledge of the
dynamics of unzipping. In practice, any functional form for ${\cal P}({\cal
N}|{\cal S})$ will be only approximate for a given experimental
setup. A possible way-out based on a learning principle is 
the following: in a first stage unzipping data corresponding to a known
sequence ($\lambda$-phage) are collected to caliber ${\cal P}$, in a second
stage predictions are made for new sequences.

Last of all, our study of fixed-force unzipping shows that bases located in  
local minima of the free-energy landscape are well predicted, while
maxima are much harder to predict. Accuracy could be greatly improved 
through an adequate force vs. time scheme capable of bringing 
the fork in the right place and making it spend time there.
Investigation of the fixed-velocity case, where the force signal is
remarkably affected by single base mutation \cite{Boc02},  
will be very interesting. 

In conclusion, we hope the present study will motivate further work
to assess and improve the performances of unzipping-based sequencing.


This work has been partially sponsored by the EC FP6 
program under contract IST-001935, EVERGROW, and the
French  ACI-DRAB \& PPF Biophysique-ENS actions.

\end{document}